\documentclass[conference]{IEEEtran}
\IEEEoverridecommandlockouts
\usepackage{cite}
\usepackage{amsmath,amssymb,amsfonts}
\usepackage{algorithmic}
\usepackage{graphicx}
\usepackage{textcomp}
\usepackage{xcolor}
\newcommand{\cambio}[1]{\textcolor{black}{#1}}

\def\BibTeX{{\rm B\kern-.05em{\sc i\kern-.025em b}\kern-.08em
    T\kern-.1667em\lower.7ex\hbox{E}\kern-.125emX}}
\begin{document}

\title{Thinking Outside the [Chat]Box: Bridging Computer Science and Industrial Design for Cognitive-Inclusive Generative AI\\
}

\author{\IEEEauthorblockN{1\textsuperscript{st} Virginia Francisco}
\IEEEauthorblockA{\textit{Facultad de Informática} \\
\textit{Universidad Complutense de Madrid}\\
Madrid, Spain \\
virginia@fdi.ucm.es}
\and
\IEEEauthorblockN{2\textsuperscript{nd} Daniel Guasch}
\IEEEauthorblockA{\textit{Universitat Politècnica de Catalunya}\\
Villanova i la Geltru, Spain \\
daniel.guasch@upc.edu}
\and
\IEEEauthorblockN{3\textsuperscript{rd} Raquel Hervás}
\IEEEauthorblockA{\textit{Instituto de Tecnología Del Conocimiento}\\
\textit{Universidad Complutense de Madrid}\\
Madrid, Spain \\
raquelhb@fdi.ucm.es}
}

\maketitle

\begin{abstract}
Current Generative AI (GenAI) interfaces remain largely constrained to chatbox interaction, which can impose high cognitive demands on users and create substantial barriers for people with intellectual disabilities (ID), including prompt formulation difficulties, response overload, and limited mechanisms to assess information reliability. To explore alternative interaction models for cognitive accessibility, we conducted a cross-disciplinary co-design challenge in which two student cohorts (Computer Science and Industrial Design) developed interface concepts from the same set of functional requirements (e.g., prompt scaffolding, structured output, GUI-based refinement, transparency, and personalization). Comparing the resulting proposals reveals both convergence on foundational requirements (notably initial calibration, proactive prompting, and direct manipulation of response fragments) and complementary contributions that outline a multi-layered support system. Computer Science teams primarily produced structural scaffolding, emphasizing predictability, navigability, and trust through mechanisms such as reliability indicators, explicit sources, and context management for long conversations. Industrial Design teams emphasized experiential scaffolding, focusing on pacing, attention guidance, multimodality, and proactive agency, including step-by-step response flows, focus modes, and assistant-like integrations. We synthesize these findings into a dual-layer scaffolding framework that expands the design space for cognitively accessible GenAI interaction beyond chat-centric models and motivates future work on expert refinement, technical feasibility, and empirical validation with users with ID.
\end{abstract}


\section{Introduction}
\label{sec:intro}
The emergence of Generative Artificial Intelligence (GenAI) has supposed a fundamental shift in how users interact with digital systems. Rather than relying on traditional command-based interfaces, interaction is increasingly moving toward intent-based outcome specification 
, where users describe goals in natural language and systems generate results accordingly.


However, the current user experience of GenAI is still in its infancy. The dominant interaction model, a simple conversational chatbox, presents significant usability challenges. This low-dimensional interaction forces users to translate complex mental models into precise prompts, often requiring “prompt engineering”, while providing limited support for exploration, refinement, and flexible interaction.
This barrier is especially difficult to overcome for people with intellectual disabilities (ID). As identified in a previous exploratory case study \cite{b2_interaccion}, users with ID face specific challenges when interacting with current GenAI tools:
\begin{itemize}
    \item Difficulties related to the question’s conceptualization.
    \item Cognitive overload when processing long responses.
    \item Challenges in verifying the reliability of the provided information.
    \item Struggles with the technical vocabulary and the lack of guidance for refining or simplifying answers.
\end{itemize}

Addressing this gap requires moving beyond incremental improvements to existing chat-based interfaces and exploring alternative design perspectives. We argue that rethinking GenAI interaction for cognitive accessibility can benefit from divergent cross-disciplinary approaches. To investigate this premise, we conducted a parallel co-design challenge involving two distinct disciplinary profiles: Computer Science students from the Universidad Complutense de Madrid (UCM), and Industrial Design students from the Universitat Politècnica de Catalunya (UPC). Each group approached the problem of redesigning GenAI interaction for users with ID from their respective disciplinary perspectives. This paper presents the results of this cross-disciplinary design exploration and analyzes how different design perspectives shape the resulting interaction paradigms. By comparing the software-centric approach with the industrial design perspective, we demonstrate how thinking outside the box (literally and metaphorically) leads to new interaction paradigms that are multimodal, proactive, and truly inclusive for users with ID.


\cambio{The rest of the paper is organized as follows. Section \ref{sec:relatedWork} reviews related work on cognitive accessibility, GenAI, and cross-disciplinary design. Section \ref{sec:redesignProcess} describes the redesign process, while Section \ref{sec:results} presents the results. Section \ref{sec:discussion} discusses the findings and Section \ref{sec:limitations} studies existing limitations. Finally, Section \ref{sec:conclusions} concludes the paper and outlines lines of future work.}


\section{Related Work}
\label{sec:relatedWork}




Designing for cognitive accessibility has gained increasing attention in Human–Computer Interaction (HCI), emphasizing the design of systems that reduce cognitive load, support comprehension, and enable users with diverse cognitive abilities to effectively engage with digital technologies. From the point of view of GenAI interaction \cite{b3_JACCES}, the emergence of GenAI interfaces introduces new challenges for cognitive accessibility, moving from structured, menu-based systems toward open-ended, prompt-based interactions that may increase cognitive load, reduce predictability, and challenge users’ ability to develop accurate mental models of system behavior \cite{b2_interaccion}.

Recent research has begun to explore these challenges, highlighting issues such as the need for effective scaffolding, clearer system feedback, and interaction designs that better support comprehension and decision-making \cite{b4_RSL}. These proposals emphasize inclusive design approaches that promote adaptability \cite{b5_Liu2024}, transparency \cite{b6_Haroon2024}, and active user involvement \cite{b7_Acosta-Vargas2024b}, often through multimodal interaction \cite{b8_Roomkham2024} and personalized experiences \cite{b9_Pierres2025} aligned with users’ cognitive needs. However, existing work often focuses on isolated design solutions or technical accessibility criteria rather than examining broader interaction models from a cognitive accessibility perspective.




Cross-disciplinary design has gained increasing attention as a strategy for addressing complex socio-technical challenges that require the integration of diverse perspectives and expertise \cite{b10_AIGC}. By bringing together different disciplinary perspectives, cross-disciplinary collaboration expands how design problems are framed and explored. Studies examining collaborations between designers, engineers, and domain experts indicate that differences in cognitive approaches, evaluation criteria, and communication styles can stimulate creativity and support the emergence of more innovative solutions by combining different perspectives \cite{b11_bridging_divides, b12_wereable}. For example, research in Explainable Artificial Intelligence (XAI) has demonstrated how cross-disciplinary engagement can help bridge gaps between technical system development and users’ mental models, emphasizing the role of participatory approaches in shaping emerging technologies \cite{b13_XAI}. Cross-disciplinary initiatives in specialized domains such as medical interface design \cite{b14_medical} or marketing \cite{b15_Visual} illustrate how bringing together expertise from different fields enables the integration of domain knowledge, design methodologies, and user-centered perspectives, contributing to more comprehensive redesign processes.

Co-design has emerged as a valuable approach for integrating multiple perspectives into design processes by encouraging collaborative ideation and exploration of future-oriented solutions, particularly in educational and research contexts involving students. Co-design workshops with students provided structured opportunities to explore emerging technologies like Multimodal Large Language Models (MLLMs), and generate future-oriented design concepts through iterative activities \cite{b16_multimodal}. Students from design and engineering fields were chosen to ensure diversity and representativeness in the development of a system called to facilitate idea convergence in interdisciplinary teams \cite{b10_AIGC}. However, to the best of our knowledge, no prior work has examined how co-design activities involving cross-disciplinary perspectives can inform design processes specifically oriented toward cognitive accessibility in AI-based interactions. \cambio{This gap motivates the present study on cross-disciplinary approaches to accessible interaction design.} 

\section{Cross-Disciplinary Redesign Process}
\label{sec:redesignProcess}
The redesign phase of GenAI interfaces was structured as a dual-track design challenge, aiming to translate the theoretical barriers identified in \cite{b2_interaccion} into functional interaction patterns. By deploying the same brief to two distinct student populations (design vs. engineering) we aimed to observe how disciplinary perspectives (product-centric vs. software-centric) could trigger different innovation paths for cognitive accessibility. In particular, design students are generally considered to express ideas through visualization and narrative scenarios, while engineering students prioritize logic and data \cite{b12_wereable}. 

\subsection{Design Requirements and Constraints}
To ensure comparability, both groups followed a common set of design requirements focused on reducing cognitive load and enhancing user agency:
\begin{enumerate}
    \item Prompt Formulation Support (Input): The system must facilitate the creation of complex prompts, allowing users to express intentions without relying solely on their writing skills or manual prompt engineering.
    \item Structured Response Delivery (Output): To prevent information overload, AI responses must be presented in a modular, structured format that is easy to scan and cognitive manageable, moving away from dense and unorganized blocks of text.
    \item Intent Refinement via GUI: The interface must use graphical elements (such as buttons, menus, or selectors) to capture and refine user intent, eliminating the need to rewrite text to correct or expand a query.
    \item Direct Output Manipulation: Users must be able to directly modify the generated output—adjusting tone, length, style, or format—through interface controls, facilitating the adaptation of information to their needs.
    \item Verification and Transparency: Mandatory mechanisms to display information reliability and direct links to primary sources must be included, allowing for straightforward verification of the provided data.
    \item Efficient Context Management: The design must enable efficient navigation, filtering, and reuse of information from long conversation threads, preventing information loss in extensive chat histories.
    \item User Profile Personalization: The system must allow users to define preferences for how they ask questions and receive information (e.g., preferred formats, level of detail, or reading pace), adapting the system to each individual's cognitive style.
\end{enumerate}

In both cohorts, we encouraged unconstrained ideation: students were asked to prioritize interaction concepts and user experience without considering technical feasibility or implementation constraints.

\subsection{Description of the Participant Groups}
Group A consisted of 3rd and 4th-year Computer Science students from the UCM ($n=35$), organized into eight working groups of 4–5 members. The challenge was conducted over one month within a structured design exercise context and included six dedicated classroom sessions. In their design work, teams emphasized predictable interaction by drawing on established UI patterns and produced high-fidelity prototypes in Figma, with a focus on screen-based interactions and system-level accessibility settings. Consistent with the course emphasis on web design, Group A was required to design a web-based interface.


Group B consisted of 4th-year Industrial Design and Product Development Engineering students from UPC ($n=15$). The participants were organized into five working groups of 3 members each. The design tasks were carried out as part of the course's practical work and were distributed across seven weekly sessions. These sessions were structured around a User-Centered Design (UCD) cycle with the following planning: work plan, specifications, conceptual design, prototyping, evaluation, and conclusions. 


\subsection{Data Collection and Analysis}

\cambio{In the case of the Computer Science cohort, each team submitted: (1) a link to the interactive Figma prototype, (2) a written report describing how the proposed interface addressed the required interaction challenges, and (3) a rationale explaining the main design decisions adopted throughout the redesign process. In addition, all teams conducted an in-class oral presentation of their proposals, where they explained the functionality of the prototype and justified the most relevant interaction and accessibility decisions. No intermediate deliverables were formally collected. However, throughout the classroom sessions, the instructor reviewed the progress of each group and provided formative feedback.}

\cambio{In the case of the Industrial Design cohort, submissions by the teams were different depending on the sessions: (1) work plan including the objectives, target users, task distribution and a timeline, and the methodology and tools to be used;
(2) specifications, defining the product's characteristics based on the understanding of the user and the desired experience; 
(3) conceptual design, with a justified product proposal demonstrating its feasibility, both technical and methodological, and from the point of view of the user; 
(4) prototype of the final product, including the prototyping process; 
(5) evaluation of the product with users, with a methodological description of the evaluation process and the obtained results; 
and (6) conclusions, along with some reasoning behind the work's evolution and the learned lessons. 
} 


\cambio{The comparative analysis between groups was conducted through qualitative inspection of the final prototypes, written rationales, and presentation materials. The authors reviewed the proposed interaction mechanisms, interface structures, and accessibility strategies to identify recurring design patterns (convergences) as well as distinctive disciplinary approaches and novel interaction proposals (divergences). Given the exploratory nature of the study, this analysis was interpretative rather than based on a formal coding scheme or a quantitative evaluation framework.}

\section{Results}
\label{sec:results}
This section presents the design solutions developed by the two groups to address the identified cognitive barriers in GenAI. Although both groups worked from the same functional requirements, their disciplinary backgrounds led to distinct complementary design paradigms. 

\subsection{Group A: Computer Science students}
\label{subsec:resultsGroupA}
The solutions developed by the Computer Science students at UCM focused on transforming the ``blank canvas'' of the chatbox into a structured and predictable workspace. 
Their main proposals addressing the seven functional requirements were as follows:
\begin{itemize}
    \item Prompt Formulation Support (Input): To reduce the anxiety generated by the ``empty box'', the students propose a real-time prompt auditing mechanism. Rather than just flagging errors, the system proactively identifies missing information in the user's request and suggests specific improvements to complete the intent as seen in Figures \ref{fig:modoConstructorGUIa} and \ref{fig:modoConstructorLUMI}. This was complemented in some cases by a visual ``prompt quality'' indicator, such as the one shown at the bottom of Figure \ref{fig:modoConstructorGUIa}, and non-intrusive orthographic suggestions that require user confirmation before prompt submission.
        \begin{figure}[htbp]
        \centering
        \includegraphics[width=0.7\columnwidth]{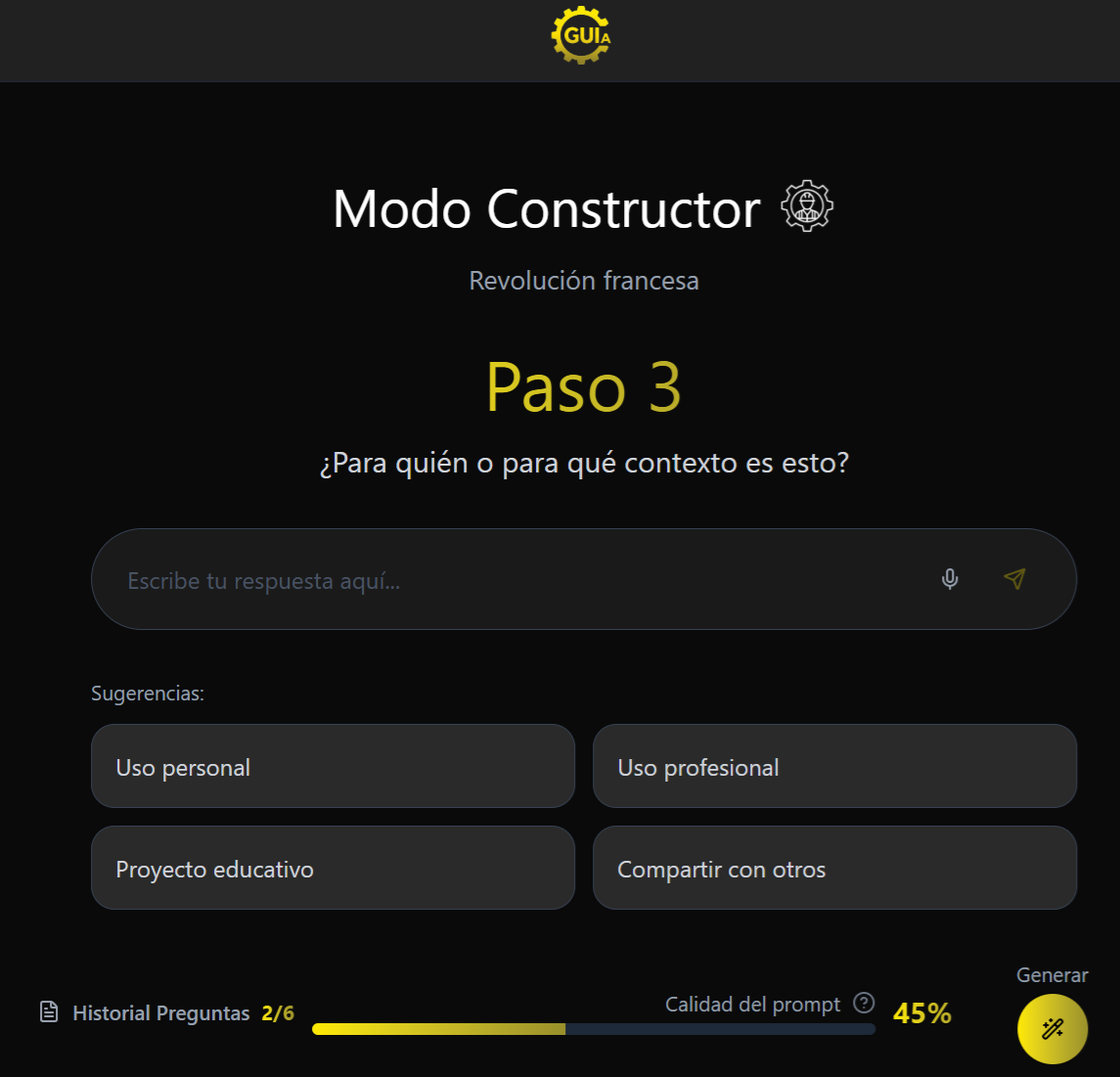}
        \caption{Real-time audit design suggested by students \cambio{from Group A} (I)}
        \label{fig:modoConstructorGUIa}
        \end{figure}

        \begin{figure}[htbp]
        \centering
        \includegraphics[width=0.95\columnwidth]{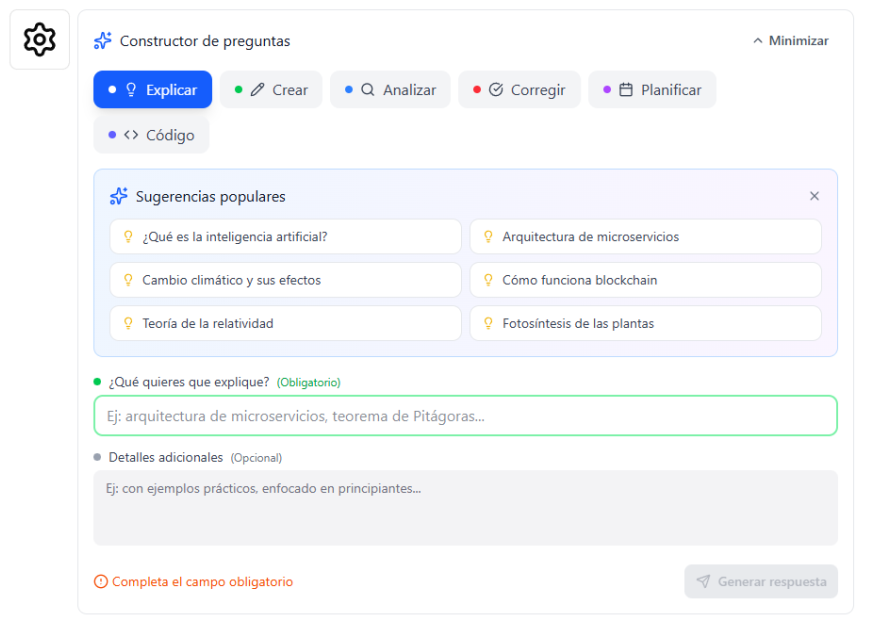}
        \caption{Real-time audit design suggested by students \cambio{from Group A} (II)}
        \label{fig:modoConstructorLUMI}
        \end{figure}
        
    \item Structured Response Delivery (Output): To prevent cognitive overload, responses were redesigned to be modular and expandable. A notable innovation was the inclusion of an automated ``Glossary Block'' at the end of each response (as shown in the purple highlighted section of Figure \ref{fig:respuestaLUMI}), which extracted and explained all technical terms, acronyms, and mnemonics in a clear manner.

        \begin{figure}[htbp]
        \centering
        \includegraphics[width=0.85\columnwidth]{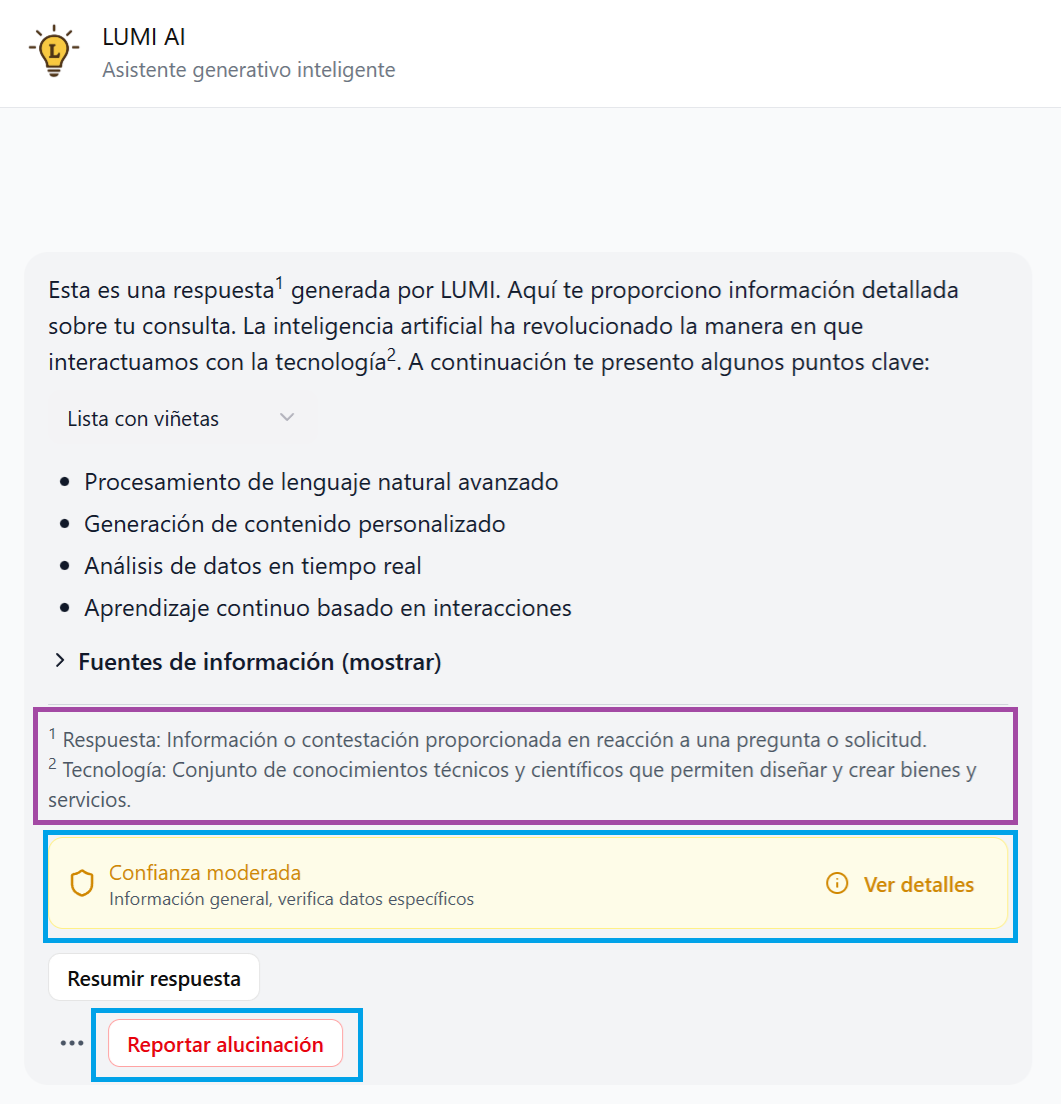}
        \caption{Response Interface with Glossary and Reliability Mechanisms \cambio{(Group A)}}
        \label{fig:respuestaLUMI}
        \end{figure}
        
    \item Intent Refinement via GUI: Several groups proposed that before starting a session, users define the AI's ``persona'' (who they want to talk to), the specific type of help needed, their own identity, mandatory restrictions, the response tone, and the desired degree of concreteness. Additionally, tools were added to provide explicit context to uploaded files, guiding the AI’s interpretation.
    \item Direct Output Manipulation: Several prototypes enabled users to interact with specific fragments of the AI's response. This included tools to save, merge, and refine response fragments. Moreover, students suggested that users could select any complex term to see an instant definition, which was then saved to a personal glossary as seen in Figure \ref{fig:glosarioGUIa}. If the term reappeared in future interactions, it would be automatically highlighted with a hover-definition (see Figure \ref{fig:respuestaGUIa}).

        \begin{figure}[htbp]
        \centering
        \includegraphics[width=0.7\columnwidth]{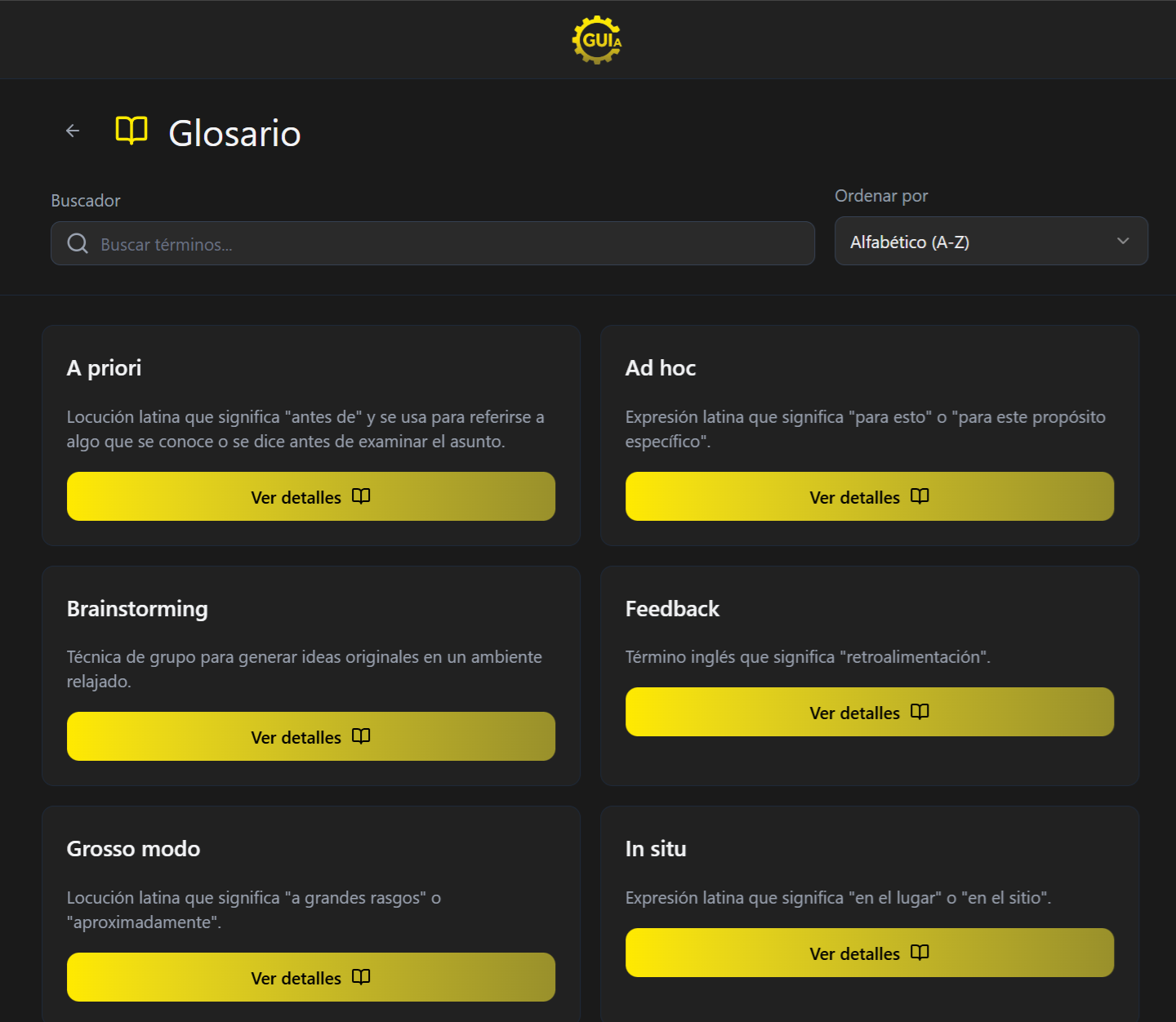}
        \caption{Personal Glossary \cambio{(Group A)}}
        \label{fig:glosarioGUIa}
        \end{figure}

        \begin{figure}[htbp]
        \centering
        \includegraphics[width=0.8\columnwidth]{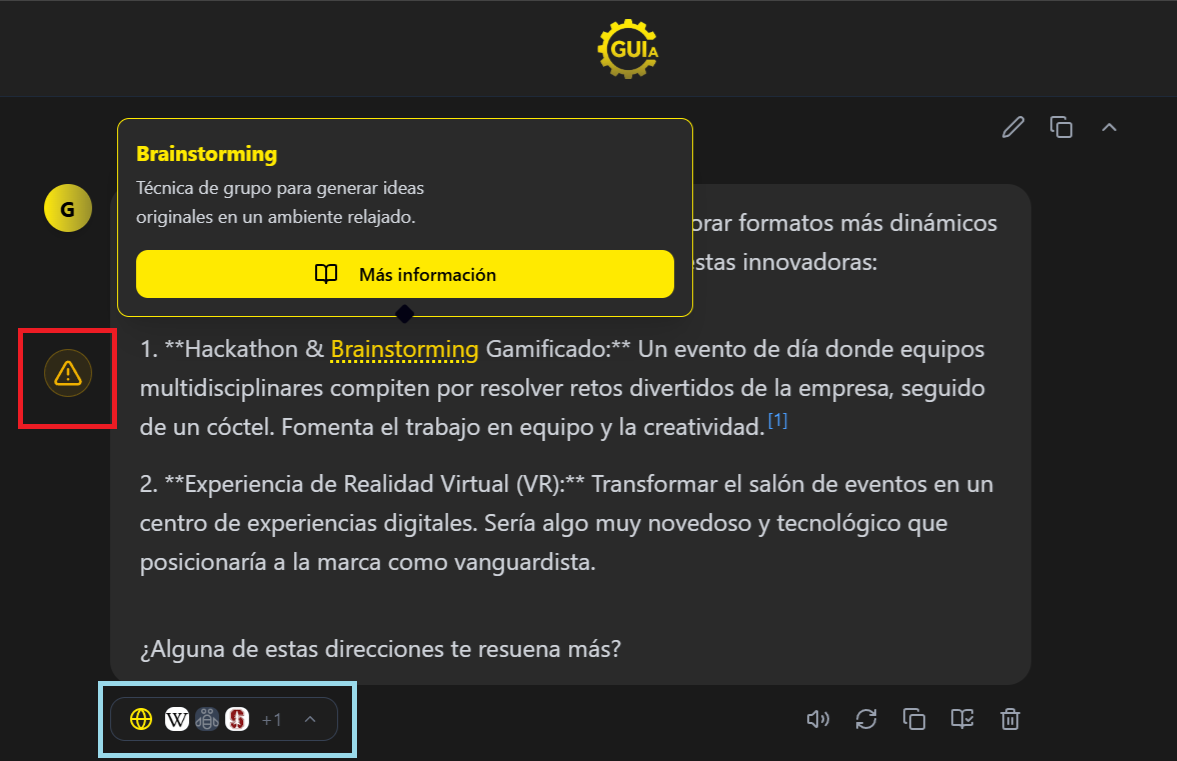}
        \caption{Highlighted Difficult Words with Hover Definitions \cambio{(Group A)}}
        \label{fig:respuestaGUIa}
        \end{figure}
        
    \item Verification and Transparency: Various groups of students developed a robust ``Reliability Traffic Light'' system (Green/Yellow/Red) (as shown in Figures \ref{fig:respuestaGUIa} -yellow indicator outlined in red- and \ref{fig:analisisConfianzaLUMI}). Beyond the overall color code, the interface shown in Figure \ref{fig:analisisConfianzaLUMI} displays a detailed breakdown of the score, including strengths, warnings, and factors analyzed. To promote verifiability and transparency, each answer is accompanied by the specific sources used by the GenAI (as shown in the section outlined in blue of Figure \ref{fig:respuestaGUIa}), and its own reliability indicator (as shown in Figure \ref{fig:fuentesGUIa}). Users can manually add or remove sources to regenerate answers as shown in Figure \ref{fig:fuentesGUIa} and utilize a dedicated button to report suspected AI hallucinations (button outlined in blue in Figure \ref{fig:respuestaLUMI}).

        \begin{figure}[htbp]
        \centering
        \includegraphics[width=0.7\columnwidth]{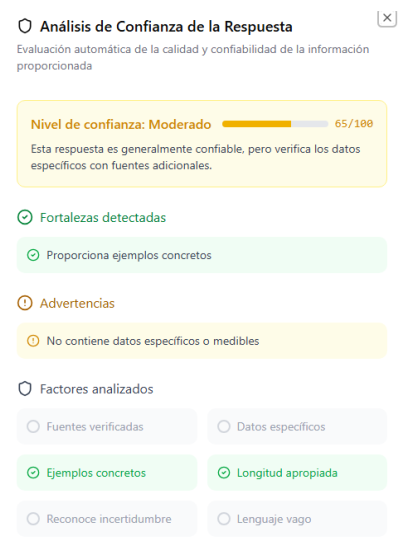}
        \caption{Confidence analysis of the proposed response \cambio{(Group A)}}
        \label{fig:analisisConfianzaLUMI}
        \end{figure}

        \begin{figure}[htbp]
        \centering
        \includegraphics[width=0.8\columnwidth]{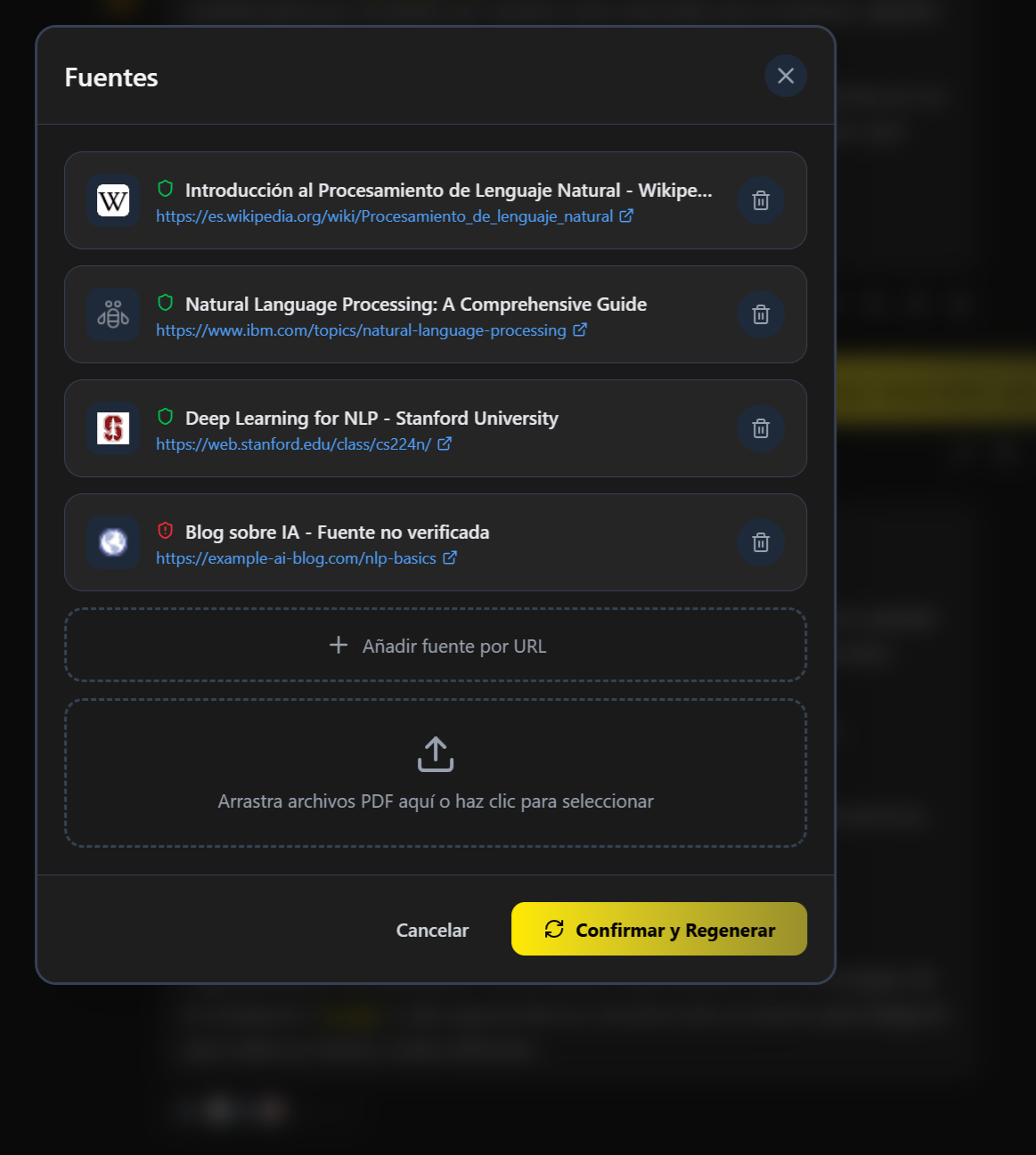}
        \caption{Analysis of the sources \cambio{(Group A)}}
        \label{fig:fuentesGUIa}
        \end{figure}
        
    \item Efficient Context Management: To manage long conversations, an emerging pattern was a two-tier organizational system: a general chat history and a specific message history within each chat. Clicking on a specific message allows the user to jump directly to that part of the conversation. This is complemented by collapsible message blocks, internal search bars, and a "Favorites" system.
    \item User Profile Personalization and Accessibility: Advanced accessibility settings were integrated, including support for Sign Language (SL) inputs and simplified UI modes. The proposed designs allowed for deep personalization of the interface layout to match each individual's cognitive style and sensory preferences, providing a high-quality help system to guide users throughout the experience.
\end{itemize}

Taken together, the proposals reveal three recurring design themes reflecting a software-centric approach: structured interaction scaffolding, transparency mechanisms, and user-controlled refinement tools.

\subsection{Group B: Industrial Design students}
\label{subsec:resultsGroupB}
The proposals developed by students in the Industrial Design and Product Development degree shifted the design approach from optimizing the digital workspace to shaping the interaction experience and the assistant's proactive nature. 
Their five designs, developed through a complete User-Centered Design cycle with real users, prioritized sensory feedback, emotional connection, and reduced interaction costs to maintain engagement without overwhelming users.

In their initial conceptual analysis, students concluded that 
various levels of “complexity” had to be considered:
\begin{itemize}
    \item  Information complexity. Information
    must be tailored to users' intellectual capabilities. Input may need to be enriched with contextual information, and responses should be simplified to facilitate comprehension.
    \item  Interface complexity. 
    Interfaces must allow for a high degree of customization, adjust the number of available options, analyze the arrangement of elements, etc.
    \item  Interaction complexity. Interaction should support a wide spectrum of users through alternative and augmentative communication systems and by enhancing the system’s perceived empathy.
\end{itemize}

A key insight emerged from analyzing how users conceptualized the interaction device. Although participants consistently preferred a smartphone, they described two distinct mental models: (1) the device as a workspace where users arrange elements that elicit a response, and (2) the device as a communication channel enabling dialogue with the GenAI system. These mental models shaped users’ expectations of the system’s role: either responding to a query (chatbot-style interaction) or actively guiding task resolution and performing actions on the user’s behalf (personal-assistant interaction). This shift also surfaced an initial adoption barrier—perceived difficulty, limited confidence, and insecurity. Several proposals addressed this barrier through avatars designed to scaffold early steps, guide communication, and provide supportive prompts. 
This group’s proposals can be summarized as follows:
\begin{itemize}
    \item Prompt Formulation Support (Input): To eliminate input paralysis, some students developed a ``Guided Mode''. Instead of an empty text field, the system initiates a simplified dialogue as shown in Figure \ref{fig:consultaPasoAPaso}, asking the user step-by-step questions to help them build their query.

        \begin{figure}[htbp]
            \centering
        
            \begin{minipage}{0.35\columnwidth}
                \centering
                \includegraphics[width=\linewidth]{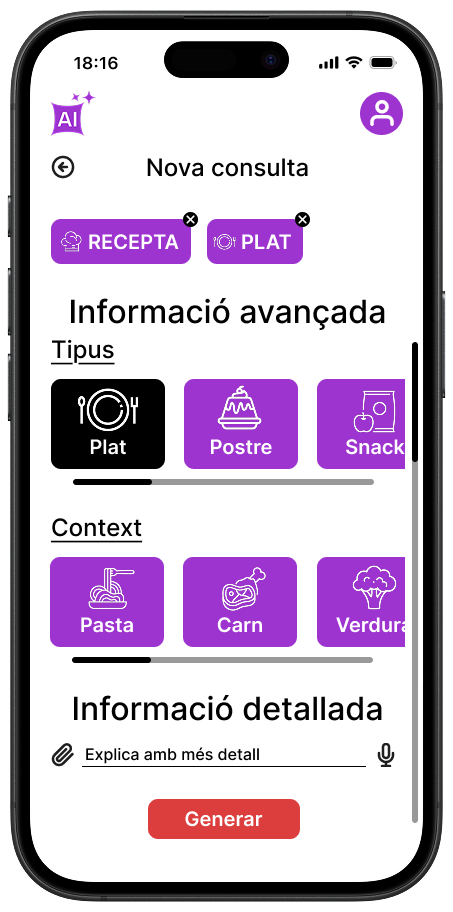}
                \caption{Guided Mode \cambio{(Group B)}}
                \label{fig:consultaPasoAPaso}
            \end{minipage}
            \hspace{0.02\columnwidth}
            \begin{minipage}{0.33\columnwidth}
                \centering
                \includegraphics[width=\linewidth]{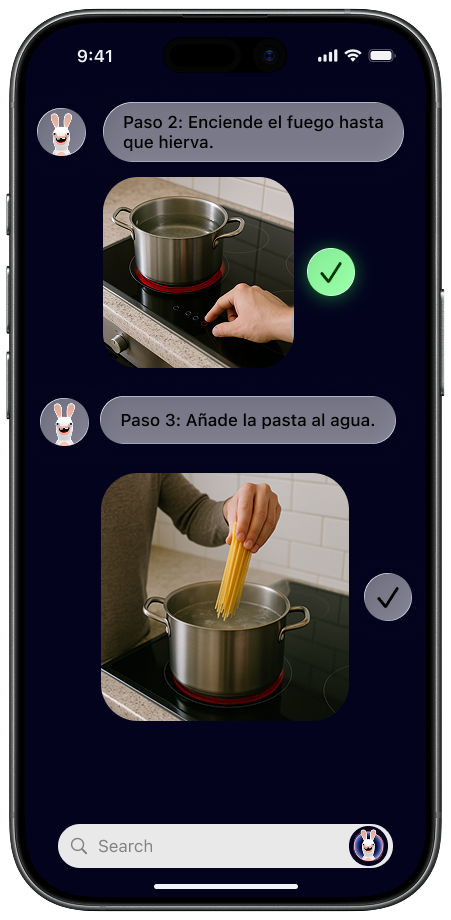}
                \caption{Proposed step-by-step sequence response \cambio{(Group B)}}
                \label{fig:pasoapasoRabbit}
            \end{minipage}
        \end{figure}

        
    \item Structured Response Delivery (Output): Two groups proposed that responses were not delivered as a whole but as a step-by-step sequence as seen in Figure \ref{fig:pasoapasoRabbit}. To ensure comprehension, the user must ``tick'' a checkbox after each segment, and the AI only reveals the next step once the previous one is confirmed. Another group proposed multiple answer formats (e.g., text or image-based) adapted to different cognitive needs. 

    
    \item Intent Refinement via GUI: As can be seen in Figure \ref{fig:manipulacionRespuesta}, a group proposed a dedicated ``Refinement Panel'' placed on the left side of the response. This panel offers quick, contextual buttons to simplify, expand or change perspective without the user having to type a new prompt.

         \begin{figure}[htbp]
            \centering
        
            \begin{minipage}{0.34\columnwidth}
                \centering
                \includegraphics[width=\linewidth]{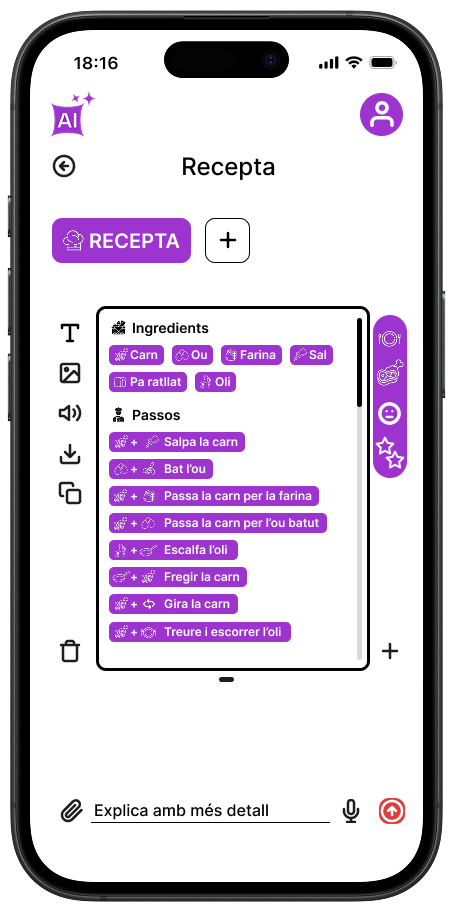}
                \caption{Dedicated Refinement Panel \cambio{(Group B)}}
                \label{fig:manipulacionRespuesta}
            \end{minipage}
            \hspace{0.02\columnwidth}
            \begin{minipage}{0.35\columnwidth}
                \centering
                \includegraphics[width=\linewidth]{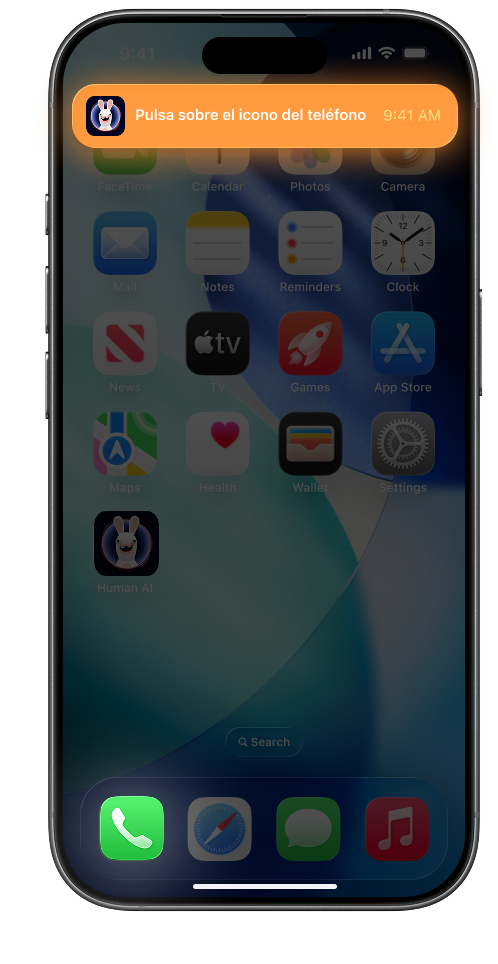}
                \caption{Focus Mode \cambio{(Group B)}}
                \label{fig:FocusMode}
            \end{minipage}
        \end{figure}
        
        
    \item Direct Output Manipulation: Users are given total control over the response personality, including the ability to adjust the intonation (e.g., more encouraging, more formal), style, and format through intuitive sliders or presets.
    \item Attention Guidance (Scaffolding): A standout feature was the ``Focus Mode'', as shown in Figure \ref{fig:FocusMode}, when the system guides a user through a task on their device, the system darkens the entire screen except for the specific element the user needs to interact with, physically guiding their attention and reducing visual noise.

        
    \item Cognitive Calibration and Personalization: A group introduced a sophisticated initial calibration phase (shown in Figure \ref{fig:configuracionInicialAI-uda}). Beyond selecting an avatar (choosing between an emotionally expressive character or a minimalist voice sensor) and adjusting font sizes, the system includes a reading speed test: the user is asked to read a short text, and the AI measures the time taken to automatically synchronize the speed at which it displays future information.

        \begin{figure}[htbp]
        \centering
        \includegraphics[width=0.8\columnwidth]{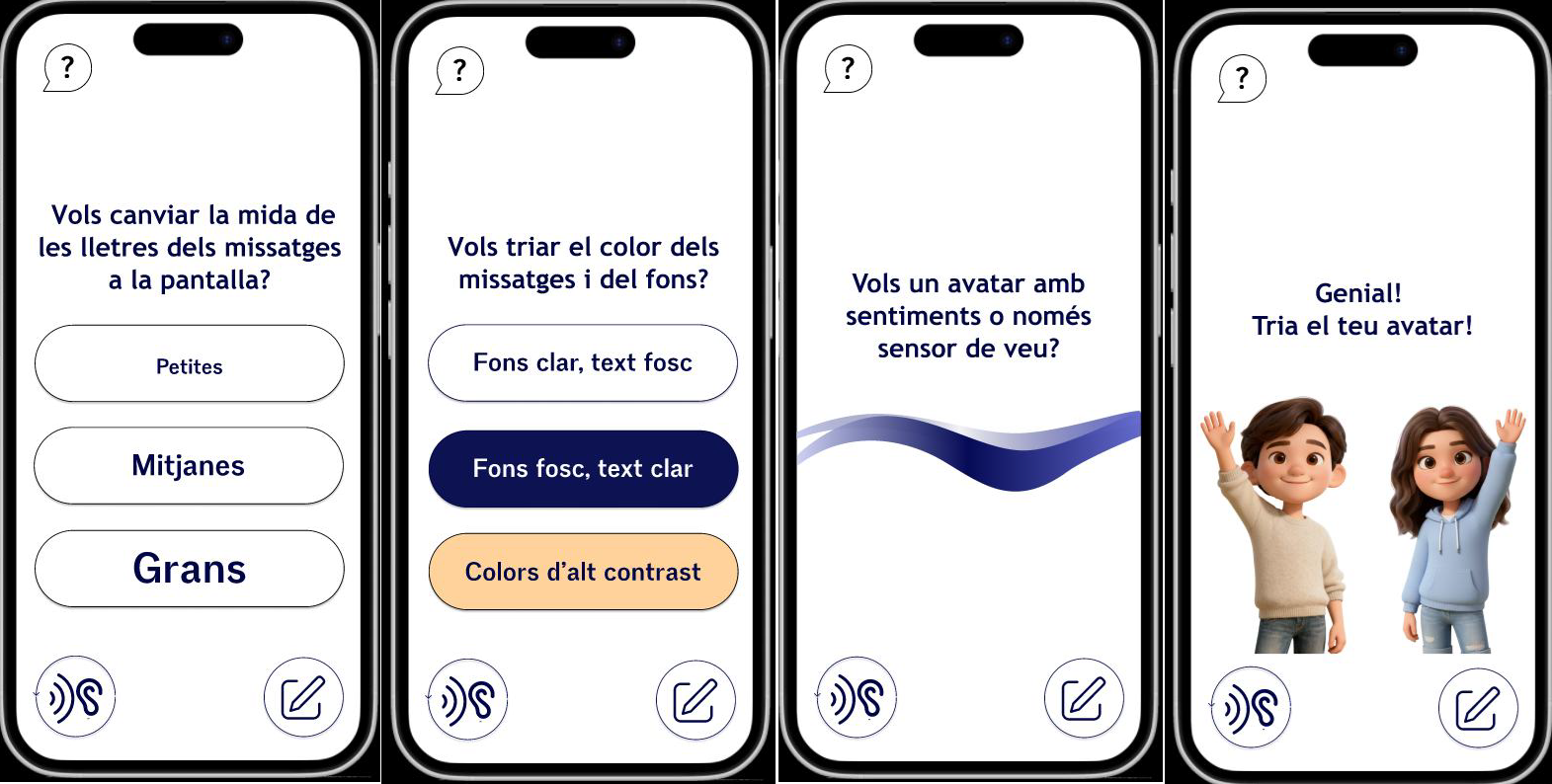}
        \caption{Initial calibration phase proposed \cambio{by Group B}} 
        \label{fig:configuracionInicialAI-uda}
        \end{figure}
        
    \item Integrated Agency and Real-world Support: Several groups proposed an integrated personal assistant embedded in the user’s operating system or physical environment (Figure \ref{fig:asistenteRabbit}). Rather than returning information in-app, the assistant was designed to support everyday tasks through proactive guidance and action, aiming to promote autonomy in real-world activities.

        \begin{figure}[htbp]
        \centering
        \includegraphics[width=0.8\columnwidth]{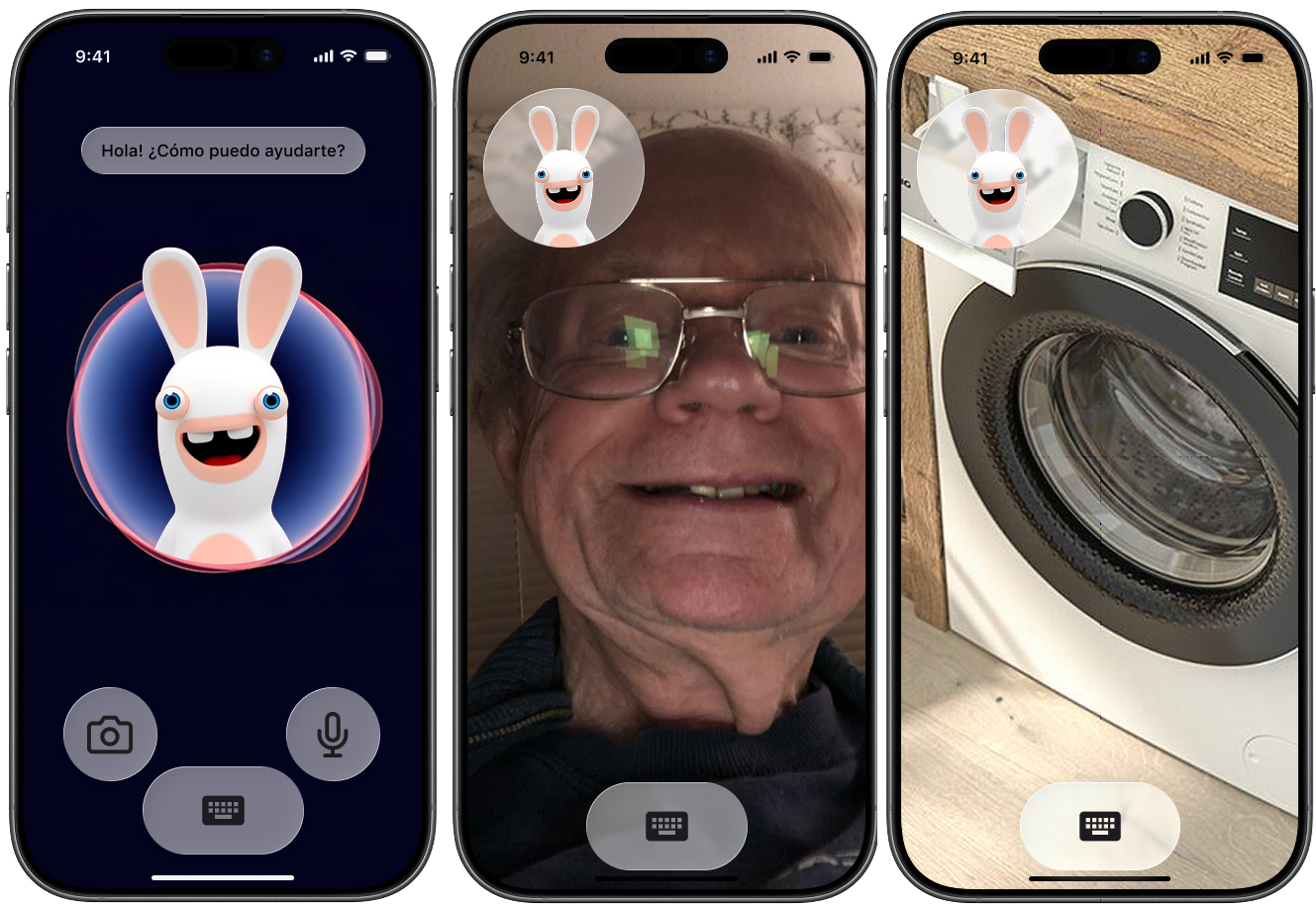}
        \caption{Integrated personal assistant \cambio{(Group B)}} 
        \label{fig:asistenteRabbit}
        \end{figure}
\end{itemize}

Collectively, these proposals shift the design paradigm toward experiential guidance, multimodal interaction, and proactive agency, highlighting a move beyond traditional chat-based interaction models.

\section{Discussion}
\label{sec:discussion}
Comparing Group A (Computer Science) and Group B (Industrial Design) reveals two complementary paradigms for cognitively accessible GenAI interaction. While their disciplinary backgrounds differ, their convergence on several key features suggests a core set of accessibility requirements, while their unique contributions outline complementary dimensions of a multi-layered support system.


Taken together, the proposals point to two complementary scaffolding paradigms: \textit{structural scaffolding}, focused on transparency, predictability, and information reliability, and \textit{experiential scaffolding}, centered on guiding user attention, pacing interaction, and reducing input complexity. Both groups converged on the need to move away from the static chatbox toward a highly customizable and guided experience. This common ground suggests that for users with ID, the following elements appear to be foundational design requirements:
\begin{itemize}
    \item The initial configuration wizard: Both groups identified that interaction should begin with a calibration. Personalizing the interface layout, sensory preferences, and the GenAI’s personality (intonation, style, and tone) was considered essential for a safe and accessible environment.
    \item Proactive Prompting: Whether through Group A's system of identifying missing information or Group B's guided mode, both approaches agree that the system must take the lead in the input phase to prevent input paralysis.
    \item Direct Manipulation of Fragments: The convergence on allowing users to interact with specific response fragments (via Group A’s fragment selection or Group B’s refinement panel) suggests that users need to refine GenAI outputs without the burden of re-typing or re-prompting.
\end{itemize}

Beyond this shared baseline, each discipline contributed a distinct scaffolding layer. Group A's proposals can be interpreted as contributing a structural scaffolding layer, addressing transparency and trust in the presence of AI hallucinations. The ``Reliability Traffic Light'' combined with explicit source citations and a button to report hallucinations provides a framework for critical thinking and verification that is often missing in accessible tools. These features operationalize structural scaffolding by making uncertainty visible, actionable, and verifiable. Group A’s two-tier organizational system (general history vs. specific message history) and collapsible blocks solve the problem of cognitive overload in long-form interactions, making the conversation navigable through search bars and favorites. The ``Glossary Block'' at the end of responses serves as an automated scaffolding tool, ensuring that technical jargon does not become a barrier to comprehension. 

In parallel, the Industrial Design students provided the experiential scaffolding, focusing on how the information is actually consumed and acted upon. A significant finding is the transition from holistic responses to a sequential check-box flow. This ensures the user is not overwhelmed and that the system waits for the user’s cognitive pace. By generating multiple versions of an answer (including images), Group B acknowledges that cognitive styles vary. The ``Focus Mode'' addresses the critical requirement of attention guidance. By darkening the screen and highlighting only the relevant task, the system physically scaffolds the user's attention, reducing environmental and digital noise.

The results suggest that inclusive GenAI may benefit from integrating both approaches. Group A provides the structural reliability (the what and the why), while Group B provides the experiential scaffolding (the how). By combining Group A’s advanced accessibility settings (like Sign Language input) with Group B’s empathic assistants, this combination could enable systems that are not only compliant with guidelines but truly proactive in their support. 

Our findings suggest that moving ``beyond prompts'' for users with ID can be framed as mixed-initiative teamwork: the system collaborates in intent formation (through guided input), supports iterative co-editing (through GUI refinement and fragment-level control), and sustains accountable interaction (through visible uncertainty and actionable verification). In this partnership model, GenAI does not merely answer requests, but helps structure tasks, manage cognitive workload, and maintain transparency throughout the interaction.


\section{Limitations}
\label{sec:limitations}

\cambio{One important limitation of this study is the asymmetry between the two educational contexts. Although both cohorts received the same high-level design challenge and functional requirements, the groups operated under different pedagogical and methodological conditions. Group A (Computer Science students) was explicitly constrained to designing a web-based interface, whereas Group B (Industrial Design students) was not restricted to a specific platform or interaction modality. This difference may have influenced the resulting proposals, favoring workspace-oriented and screen-centric solutions in Group A, while enabling more exploratory and multimodal concepts in Group B.}

\cambio{Additional asymmetries include differences in cohort size (35 vs. 15 students), number of working sessions (six vs. seven), team organization, and course structure. In particular, Group B followed a complete User-Centered Design cycle integrated into the course methodology, while Group A approached the challenge primarily as a structured interface redesign exercise. These contextual differences limit the extent to which direct comparisons between groups can be interpreted as purely disciplinary differences.}

\cambio{Another limitation is that the identification of convergences and divergences relied on qualitative inspection of the prototypes, reports, and presentations conducted by the authors. Consequently, the findings should be interpreted as emerging design tendencies rather than statistically generalizable results.}

\cambio{Finally, users with ID did not directly participate in the design activities reported in this paper. While the redesign requirements were grounded in barriers identified in prior work with users with ID, the concepts presented here reflect the perspectives of students and researchers rather than co-designed solutions developed together with the target population. This absence limits the validity of the proposals and means that the accessibility benefits discussed in this work remain hypothetical until validated through participatory design and empirical evaluation with users with ID.}


\section{Conclusions and Future Work}
\label{sec:conclusions}
This study suggests that the cognitive gap in current GenAI systems is not an inherent limitation of the technology, but a design challenge that can be bridged through cross-disciplinary innovation. By deploying a set of seven functional requirements to both Computer Science and Industrial Design students, we have identified a complementary set of solutions that reframe how inclusive GenAI interfaces may be conceptualized.

The main contribution of this work is the identification of a multidimensional framework for inclusive GenAI. Our findings indicate that technical transparency (integrity and reliability) must be inseparable from interactional scaffolding (pace and attention guidance). The convergence of Computer Science and Industrial Design perspectives suggests that the future of inclusive GenAI lies in systems that are not only ``easy to read'' but ``easy to navigate'', providing a structured environment where the user remains in control without being overwhelmed by the complexity of the underlying technology. We refer to this as a dual-layer scaffolding framework, integrating structural reliability (transparency, verification, and navigability) with experiential guidance (attention management, pacing, and multimodal interaction).

To transition these innovative concepts into real-world applications, future research will follow a four-track roadmap:
\begin{itemize}
    \item Professional Refinement and Expert Audit: The solutions proposed by the students must undergo a rigorous review process. We will involve HCI experts to benchmark these ideas against professional standards, grounding the students’ creativity into scalable, industry-standard design systems. This audit will specifically focus on correcting accessibility pitfalls, such as replacing abstract percentages (used to indicate prompt quality or reliability) with more intuitive, non-numerical visual metaphors that better suit the cognitive needs of the final users.
    \item Co-Design and Participatory Research: We aim to move from a ``design-for'' to a ``design-with'' approach. We will conduct new co-design workshops involving both HCI experts and individuals with ID. This process will allow users to provide their own design solutions and perspectives, ensuring that future iterations of the interface truly reflect their lived experiences and unmet needs.
    \item Technical Feasibility and Prototyping: Many of the most disruptive proposals face significant technical barriers. Real-time hallucination detection for the ``Reliability Traffic Light'' remains an open challenge in LLM research due to the ``black box'' nature of these models. Similarly, features like ``Focus Mode'' would require deep operating system integration (APIs) that are currently restricted. We plan to develop proof-of-concept versions to test their viability within current technical constraints.
    \item Empirical Validation: The consolidated prototype resulting from the expert audit and co-design sessions will undergo longitudinal testing with the target users. This will measure the impact of our design features on user autonomy and task completion in real-world scenarios.
\end{itemize}

By bridging the gap between student innovation, professional expertise, and users with ID participation, we can ensure that GenAI becomes a vehicle for cognitive inclusion.

\section*{Acknowledgment}
The authors would like to thank all participating students for their creativity, commitment, and thoughtful engagement throughout the design process.

This publication is part of the R\&D\&I project HumanAI-UI (Grant No. PID2023-148577OB-C22) from the Ministerio de Ciencia, Innovación y Universidades (Spanish Ministry of Science, Innovation and Universities).


\vspace{12pt}


\begin{thebibliography}{00}

\bibitem{b2_interaccion} V. Francisco, R. Hervás, and R. García-Mata, ``Challenges in accessing generative AI for users with cognitive disabilities: an exploratory case study,'' Proceedings of the XXV International Conference on Human-Computer Interaction (INTERACCIÓN 2025), URL: https://ceur-ws.org/Vol-4000/paper09.pdf
\bibitem{b3_JACCES} D. Guasch, C. Rodrigo, V. Francisco, and R. Hervás, ``Roadmap to inclusive Artificial Intelligence for persons with intellectual disability,'' Journal of Accessibility and Design for All, 15(2), 74-96, 2025. https://doi.org/10.17411/jacces.v15i2.625
\bibitem{b4_RSL} V. Francisco, C. Rodrigo, F. Iniesto, and R. Hervás, ``Cognitive Accessibility in Generative AI Interfaces: A Systematic Review,'' International Journal of Human–Computer Interaction, 1–17, 2026. https://doi.org/10.1080/10447318.2026.2618562
\bibitem{b5_Liu2024} X. Liu, et al., ``Understanding students’ perspectives, practices, and challenges of designing with AI in special schools,'' Proceedings of the Eleventh International Symposium of Chinese CHI, CHCHI ‘23, New York, NY, USA (pp. 197–209). Association for Computing Machinery, 2024. https://doi.org/10.1145/3629606.3629625
\bibitem{b6_Haroon2024} R. Haroon, and F. Dogar, ``TwIPS: A large language model powered texting application to simplify conversational nuances for autistic users,'' Proceedings of the 26th International ACM SIGACCESS Conference on Computers and Accessibility, ASSETS ‘24, New York, NY, USA (pp. 24:1–24:18). Association for Computing Machinery, 2024. https://doi.org/10.1145/3663548.3675633
\bibitem{b7_Acosta-Vargas2024b} P. Acosta-Vargas, B. Salvador-Acosta, S. Novillo-Villegas, D. Sarantis, and L. Salvador-Ullauri, ``Generative artificial intelligence and web accessibility: Towards an inclusive and sustainable future,'' Emerging Science Journal, 8(4), 1602–1621, 2024. https://doi.org/10.28991/ESJ-2024-08-04-021
\bibitem{b8_Roomkham2024} S. Roomkham, and L. Sitbon, ``Restarting the conversation about conversational search: Exploring new possibilities for multimodal and collaborative systems with people with intellectual disability,'' Proceedings of the 2024 Conference on Human Information Interaction and Retrieval, CHIIR ‘24, New York, NY, USA (pp. 231–242). Association for Computing Machinery, 2024. https://doi.org/10.1145/3627508.3638339
\bibitem{b9_Pierres2025} O. Pierrès, A. Darvishy, and M. Christen, ``Exploring the role of generative AI in higher education: Semi-structured interviews with students with disabilities,'' Education and Information Technologies, 30(7), 8923–8952, 2025. https://doi.org/10.1007/s10639-024-13134-8

\bibitem{b10_AIGC} J.R. Li, H.Y. Huang, T.W. Chang, C.C. Shih and H.T. Chien, ``Constructing a Cross-Disciplinary Idea Convergence System Using AIGC: A Case Study of Engineering and Design,'' 27th International Conference Information Visualisation (IV), Tampere, Finland, 2023, pp. 352-357, doi: 10.1109/IV60283.2023.00066.
\bibitem{b11_bridging_divides} C. Wang, and F. Zhang, ``Bridging Divides: A Cross-Disciplinary Design Innovation Approach to Mitigating Extreme Behavioral Expressions of Prejudice,'' Big.D, 2(1), 66–72, 2025.
\bibitem{b12_wereable} T.W. Chang, et al., ``A Wearable Emotion Display Device for Cross-Disciplinary Collaboration,'' In: Luo, Y. (eds) Cooperative Design, Visualization, and Engineering. CDVE 2025. Lecture Notes in Computer Science, vol 16092, 2026. Springer, Cham. https://doi.org/10.1007/978-3-032-06952-8\_21
\bibitem{b13_XAI} H. Sheridan, E. Murphy, D. O’Sullivan, ``Exploring Mental Models for Explainable Artificial Intelligence: Engaging Cross-disciplinary Teams Using a Design Thinking Approach,'' In: Degen, H., Ntoa, S. (eds) Artificial Intelligence in HCI. HCII 2023. Lecture Notes in Computer Science, vol 14050, 2023. Springer, Cham. https://doi.org/10.1007/978-3-031-35891-3\_21
\bibitem{b14_medical} M. Zolotova, A. Kablova, ``Exploring Cross-Disciplinary Design Dialogues: A Case-Study Workshop on Designing Medical Interfaces and the Integration of the ‘Check’ Methodological Canvas,'' In: Rau, PL.P. (eds) Cross-Cultural Design. HCII 2024. Lecture Notes in Computer Science, vol 14699, 2024. Springer, Cham. https://doi.org/10.1007/978-3-031-60898-8\_12
\bibitem{b15_Visual} H.C. Hsiao, M.D. Shieh, Y.T. Hsiao, Z.J. Jiang and J.F. Chang, ``Enhancing Cross-disciplinary Design Competence through an Innovative Visual Identity Model and Structured Teaching,'' 2025 11th International Conference on Education and Training Technologies (ICETT), Macao, China, 2025, pp. 306-311, doi: 10.1109/ICETT66247.2025.11137089.
\bibitem{b16_multimodal} P. Prasad, R. Balse, and D. Balchandani, ``Exploring Multimodal Generative AI for Education through Co-design Workshops with Students,'' In Proceedings of the 2025 CHI Conference on Human Factors in Computing Systems (CHI '25). Association for Computing Machinery, New York, NY, USA, 1–17, 2025. https://doi.org/10.1145/3706598.3714146



\end{thebibliography}
\end{document}